# Nanoscale determination of the metal-insulator transition in intercalated bulk VSe$_2$

*Wanru Ma[1], Ye Yang[2], Zuowei Liang[1], Ping Wu[1], Fanbao Meng[1], Zhenyu Wang[1]\* and Xianhui Chen[1,3,4]\**

[1]Department of Physics, University of Science and Technology of China, Hefei, Anhui 230026, China

[2]Department of Physics, Hefei University of Technology, Hefei, Anhui 230009, China

[3]CAS Center for Excellence in Quantum Information and Quantum Physics, Hefei, Anhui 230026, China

[4]Collaborative Innovation Center of Advanced Microstructures, Nanjing 210093, China

\*Correspondence and requests for materials should be addressed to Z.W. (zywang2@ustc.edu.cn) or X.-H.C. (chenxh@ustc.edu.cn).

**ABSTRACT**

Two-dimensional (2D) materials provide unique opportunities to realize emergent phenomena by reducing dimensionality. Using scanning tunneling microscopy combined with first-principles calculations, we determine an intriguing case of a metal-insulator transition (MIT) in a bulk compound, (TBA)$_{0.3}$VSe$_2$. Atomic-scale imaging reveals that the initial 4$a_0$×4$a_0$ charge density wave (CDW) order in 1T-VSe$_2$ transforms to $\sqrt{7}a_0 \times \sqrt{3}a_0$ ordering upon intercalation, which is associated with an insulating gap with a magnitude of up to approximately 115 meV. Our calculations reveal that this energy gap is highly tunable through electron doping introduced by the intercalant. Moreover, the robustness of the $\sqrt{7}a_0 \times \sqrt{3}a_0$ CDW order against the Lifshitz transition points to the key role of electron-phonon interactions in stabilizing the CDW state. Our work clarifies a rare example of a CDW-driven MIT in quasi-2D materials and establishes cation intercalation as an effective pathway for tuning both the dimensionality and the carrier concentration without inducing strain or disorder.

Two-dimensional (2D) materials such as transition-metal dichalcogenides (TMDs) provide promising platforms for exploring novel quantum states that are strongly dependent on their inherent dimensionality[1-4]. At the monolayer limit, TMDs can exhibit distinct phenomena from their bulk counterparts, including the realization of Ising superconductivity[5], the quantum spin Hall effect[6-9], and diverse charge density wave (CDW) orders[10-14]. Owing to the strong in-plane bonding and weak out-of-plane interactions, reducing the dimensionality of TMDs can generally be achieved by means of exfoliation or thin film growth. However, obtaining large-scale, stable atomically thin layers remains challenging, and additional complexities, such as external strains, might be introduced in the fabrication process. Another effective pathway to weaken the interlayer interactions is to add spacer layers or intercalate large ions into the van der Waals gap of the bulk crystal[15-23]. For example, a significant increase in the superconducting transition temperature up to 40 K has been observed in organic-ion intercalated FeSe, with a well-defined pseudogap feature above $T_c$ indicating 2D superconducting fluctuations[17], akin to the monolayer FeSe on $SrTiO_3$[24-27]. Moreover, intercalating large organic cations has been reported to give rise to Ising superconductivity in bulk $NbSe_2$[10] and to manipulate the ferromagnetic order in intercalated $Cr_2Ge_2Te_6$[16]. In this context, exploring how the ground states of TMDs change upon intercalation is of fundamental interest.

An interesting feature of TMDs is the formation of CDW orders, a collective phenomenon accompanied by the redistribution of electron density and an associated periodic lattice distortion. Despite intensive study, the underlying drive mechanism remains far from fully understood. Peierls predicted that a one-dimensional (1D) metal is unstable and prone to the formation of a CDW whose periodicity is related to the Fermi vectors of $2k_F$ and that perfect nesting in 1D opens a gap at the Fermi level and leads to a metal-insulator transition (MIT)[28]. The nesting condition is usually imperfect in 2D because of the complicated Fermi surface (FS) geometry, which generally results in a partial opening of the CDW gap depending on the nesting vectors. On the other hand, the role of electron-phonon coupling *versus* FS nesting has been recently highlighted in the formation of CDW in 2D materials[29,30]. As such, $VSe_2$ is of special interest because of the rich CDW configurations it hosts. Bulk 1T-VSe$_2$ is known to undergo a CDW transition at approximately 110 K, forming a three-dimensional (3D) superlattice of $4a_0 \times 4a_0 \times 3c_0$[31]. Reducing the dimensionality not only alters the CDW transition temperature ($T_{CDW}$)[32,33] but also leads to dramatic changes in the CDW order and electronic structure at the monolayer limit[34-38]. To date, various types of CDW orders have been reported in monolayer or few-layer VSe$_2$ films grown by molecular beam epitaxy, including $4\times4$[34,37], $\sqrt{7}\times\sqrt{3}$[34-38], $2\times\sqrt{3}$[35], $4\times\sqrt{3}$[39], and charge-strip order[35,37], whose origins have been attributed to either FS nesting[36,40] or electron-phonon coupling[34,41]. Furthermore, how these CDW orders alter the electronic structure remains highly debated[34-36], with complications arising from charge transfer, strain of the substrate, the heterointerfaces therein and even competing ferromagnetism[42]. Recently, it has been reported that the Na atoms deposited on the surface can form intercalated subsurface islands that modify the CDW configuration from the $4\times4$ phase to the $\sqrt{7}\times\sqrt{3}$ order for the top layer of VSe$_2$[43]. Therefore, intercalation could provide new insights into understanding the multiple CDW instabilities within this system.

In this work, using a combination of scanning tunneling microscopy (STM) and first-principles calculations, we present a detailed study on the emergent CDW order that is associated with an MIT in VSe$_2$ after intercalation with organic cations (TBA)$^+$. Atomic structure measurements successfully reveal a $\sqrt{7}\times\sqrt{3}$ CDW order of the VSe$_2$ layers after intercalation, in agreement with the major observations in monolayer films. In contrast to the partially opened CDW gap in previous reports[37,42-44], we observe a hard insulating gap of up to 115 meV in the single-particle spectrum, indicating that the entire FS has been gapped out at low temperatures, which leads to an MIT. Our calculations show that this energy gap is linked to the formation of the $\sqrt{7}\times\sqrt{3}$ CDW order, the magnitude of which can be further controlled by electron doping of the intercalated cations. We find that the $\sqrt{7}\times\sqrt{3}$ order is still robust in our intercalated samples where electron doping drives a Lifshitz transition, suggesting that the electron-phonon coupling should be dominant in stabilizing the CDW.

The large size of the intercalated organic cations results in a dramatic expansion of the interlayer spacing between VSe$_2$ planes, from 6.11 Å in bulk 1T-VSe$_2$ to 18.62 Å, as shown in Fig. 1A. The temperature-dependent in-plane resistivities were measured for both pristine bulk VSe$_2$ and intercalated VSe$_2$. As shown in Fig. 1B, pristine VSe$_2$ shows metallic behavior with an anomaly at approximately 110 K, which corresponds to the 4×4×3 CDW transition[31]. In contrast, the in-plane resistivity of intercalated VSe$_2$ exhibits a compelling MIT at approximately 150 K (Fig. 1C), and the hysteresis loop suggests that it is a first-order transition. Our goal is to clarify the nature of this MIT and investigate how intercalation modifies the atomic and electronic structures of VSe$_2$.

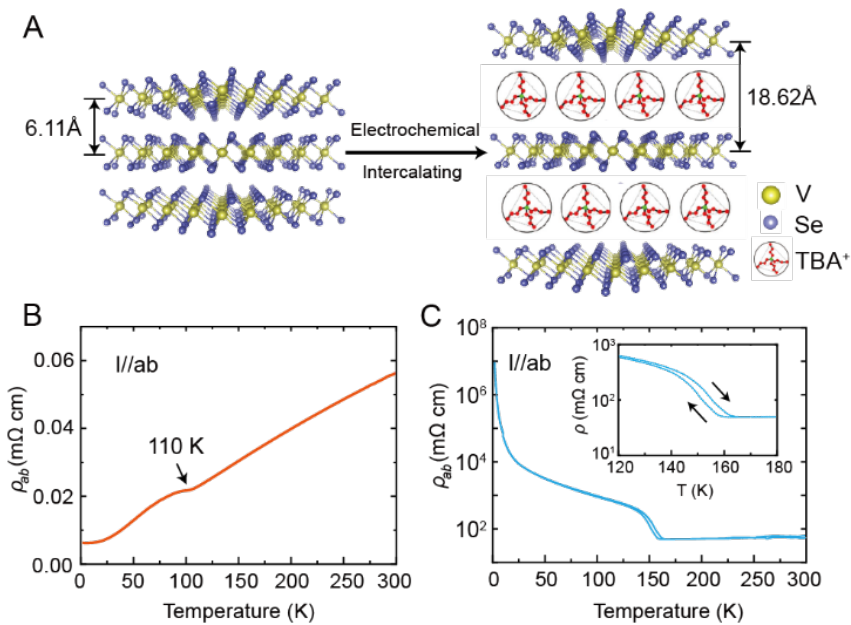

**Figure 1.** (A) Schematic illustration of the interlayer spacing changing by the electrochemical intercalation of organic cations (TBA)$^+$. (B) Temperature dependence of the in-plane resistivity of bulk 1T-VSe$_2$. (C) Temperature dependence of the in-plane resistivity for intercalated VSe$_2$. The inset shows an enlarged view of the curve around the MIT. The transition temperature is extracted from the mean value of the middle transition points in the cooling and warming curves.

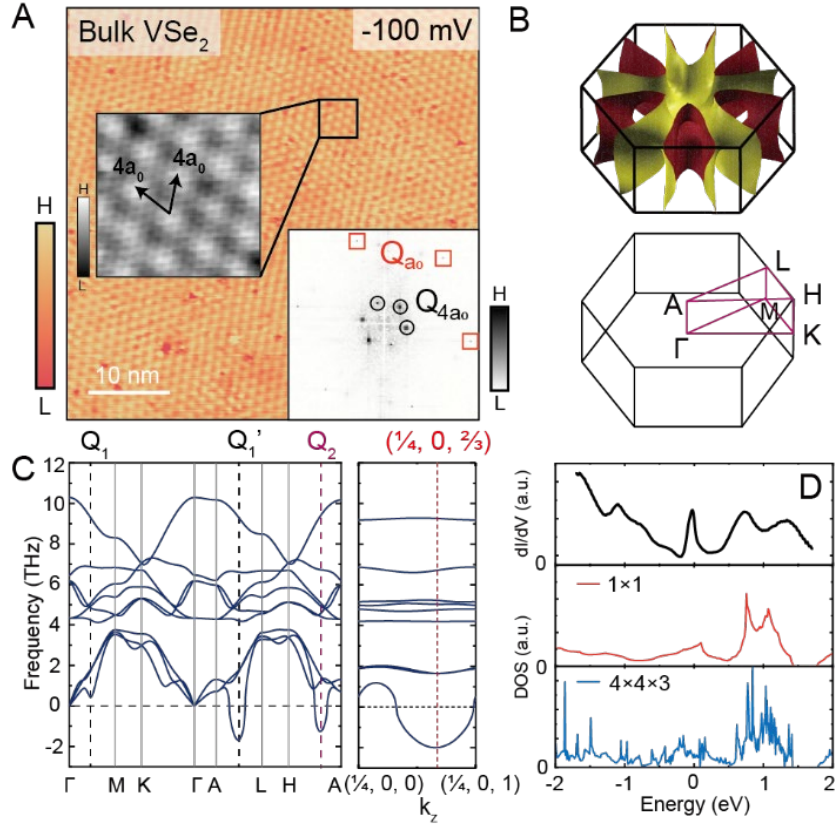

**Figure 2.** (A) Atomic-resolution STM topography of pristine VSe$_2$. The zoom-in image shows a 4×4 CDW pattern, and the inset shows its Fourier transform. The orange squares and black circles denote the atomic Bragg peaks and the 4×4 CDW peaks, respectively ($V_s$ = -100 mV, $I_t$ = 600 pA). (B) Fermi surface of pristine VSe$_2$. The bottom panel outlines the high-symmetry paths. (C) Calculated phonon dispersion of pristine VSe$_2$. Q$_1$' and Q$_2$ represent the prominent imaginary frequencies. The right panel shows the phonon dispersion along the k$_z$ line (1/4, 0, k$_z$). The major imaginary frequency is located at k$_z$ = 2/3. (D) Typical *dI/dV* spectrum for bulk 1T-VSe$_2$. The lower panel shows the calculated DOS for the original unit cell and the 4×4×3 CDW order for comparison ($V_S$ = 1.7 V, $I_t$ = 800 pA, $V_m$ = 10 mV).

First, we establish a benchmark for comparison by studying the electronic structure of pristine bulk VSe$_2$. The atomic-resolution STM topography of pristine VSe$_2$ shows a pronounced, long-range 4×4 CDW order superimposed on the atomic modulations (Fig. 2A); correspondingly, its Fourier transform (inset) also shows additional peaks at a quarter of the reciprocal-lattice vectors (Q$_{a_0}$), which we label Q$_{4a_0}$. To gain insight into the origin of this 3D CDW order, we calculated the phonon dispersion for bulk 1T-VSe$_2$. The results are presented in Fig. 2C, with high-symmetry points of the 3D Brillouin zone shown in Fig. 2B. Notably, two remarkable soft phonon modes

(negative frequencies) appear along the A-L and H-A paths, labeled as $Q_1'$ and $Q_2$. The $Q_1'$ mode corresponds to an in-plane projection of $4a_0$ modulation, whereas the $Q_2$ wavevector is related to $4\times\sqrt{3}$ distortions[39,45] and is energetically less favored here. In addition, the imaginary frequency along the $k_z$ direction, corresponding to electronically driven instabilities toward out-of-plane distortion, is dominated by a soft mode at $k_z = 2/3$. This results in a wavevector of $(1/4, 0, 2/3)$, which leads to a $4\times4\times3$ CDW distortion when the equivalent wavevectors related to lattice symmetry are considered. Therefore, the electron-phonon coupling plays a critical role in the CDW formation of bulk 1T-VSe$_2$, which is in line with a recent nonperturbative anharmonic phonon calculation[46]. The measured differential conductance ($dI/dV$) spectrum, which reflects the electronic density of states (DOS), is shown in Fig. 2D. An obvious peak-like feature is observed near the Fermi level without notable gap opening, which is consistent with the metallic nature of the pristine samples. For direct comparison, the calculated DOSs for the original unit cell and the $4\times4\times3$ CDW order are shown in the lower panel of Fig. 2D, and one can find that both match the experimental data reasonably well. This suggests that the $4\times4\times3$ distortion is rather weak compared to the atomic potential and does not significantly alter the overall electronic structure of bulk VSe$_2$.

We next turn to the intercalated samples. Performing STM measurements on these organic-VSe$_2$ hybrid crystals is a challenging task, as the randomly distributed, highly insulating organic cations (see Figure S1) usually cover the surface, making the VSe$_2$ layers difficult to access. With cryogenic temperature cleaving, we obtained several sets of atom-resolved STM data on exposed VSe$_2$ surfaces. One example is shown in Fig. 3A, where a new CDW pattern of $\sqrt{7}\times\sqrt{3}$ can be observed. The new periodicity is marked by the green and orange arrows, and a domain wall exists in the middle of this field of view. This CDW pattern aligns with the major one found in monolayer VSe$_2$ thin films[35-38,42,44] and breaks the symmetry of the lattice. The 2D Brillouin zones for the $1\times1$ normal phase and the $\sqrt{7}\times\sqrt{3}$ CDW phase are displayed in Fig. 3C, and it is clear that the Fourier transform of Fig. 3A exhibits peaks at both the Bragg vectors and the $\sqrt{7}\times\sqrt{3}$ wave vectors (Fig. 3B). Interestingly, with the $\sqrt{7}\times\sqrt{3}$ CDW order, we find an unexpected hard energy gap of up to ~115 meV that spans the Fermi level (Fig. 3D; determined by taking the logarithm of *dI/dV*), indicating an insulating ground state at low temperature. This large insulating gap is consistent with the MIT revealed by our transport data, but differs from the common observation of a slight suppression of the DOS (corresponding to a partial gap) in VSe$_2$ monolayer films[37,42,44]. Notably, the emergence of the √7×√3 order and the insulating gap has been coherently repeated at several locations separated at the macroscopic scale and should be representative of the sample (Figure S2). On the other hand, the TBA molecules exhibit a rather disordered distribution on the surface, and we have never found √7×√3 reconstructions of these molecules.

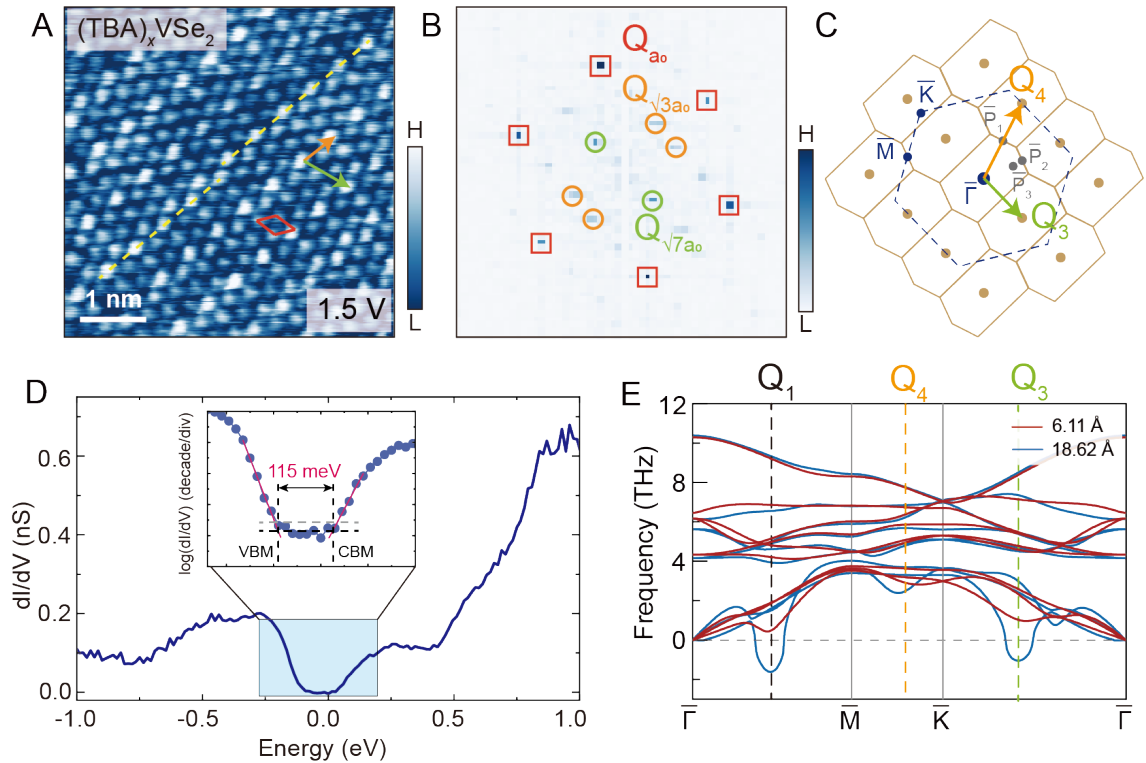

**Figure 3.** (A) Atomically resolved STM topograph of the exposed VSe$_2$ surface on the organic-VSe$_2$ hybrid crystals. The red diamond denotes the VSe$_2$ unit cell. The CDW unit cell is indicated by orange and green arrows ($V_S$ = 1.5 V, $I_t$ = 350 pA). (B) Fourier transform of the STM topography in (A). The red squares mark the atomic Bragg vectors. The orange and green circles indicate the CDW wavevectors. (C) The 2D Brillouin zones for the 1×1 cell (blue dashed lines) and the √7×√3 supercell (brown solid lines). The green and orange arrows indicate the CDW wavevectors of √7$a_0$ and √3$a_0$, respectively. (D) Typical *dI/dV* spectrum of the exposed VSe$_2$ surface. The inset shows an enlarged view of log(*dI/dV*), in which black and gray dashed lines indicate the mean of the signal ($C_{g,av}$) within the gap and two standard deviations above $C_{g,av}$ ($V_S$ = 1.5 V, $I_t$ = 500 pA, $V_m$ = 10 mV). (E) Phonon dispersion for pristine and monolayer VSe$_2$. The competing CDW wavevectors, $Q_1$ and $Q_{3,4}$, are associated with the 4×4 and $\sqrt{7}\times\sqrt{3}$ CDW distortions, with $Q_1 = 1/2\overline{\Gamma M} = 0.25b_1^*$ and $Q_3 = 3/5\overline{\Gamma K} = 0.2(b_1^* + b_2^*)$, where $b_1^*$ and $b_2^*$ are the reciprocal vectors of the 1×1 cell.

The observed $\sqrt{7}\times\sqrt{3}$ order after intercalation is highly unusual in two aspects: first, its configuration and symmetry are highly unexpected because the CDW modulation in many TMDs can usually persist even down to the monolayer limit owing to their inherently 2D nature [11,47-49]; second, the emergence of a large insulating gap via CDW ordering is very rare in 2D or bulk materials and deserves further study.

The first key role that the intercalated cations play is to reduce the dimensionality by significantly increasing the inter-VSe$_2$-layer spacing from 6.11 Å to 18.62 Å. To reveal the possible origin of the $\sqrt{7}\times\sqrt{3}$ CDW order after intercalation, we calculate the phonon dispersion of VSe$_2$ with an expanded vacuum layer. As shown in Fig. 3E, with increasing interlayer spacing, two prominent instabilities emerge in the calculated phonon bands with wavevectors of $Q_1 = 1/2 \overline{\Gamma M} = 0.25b_1^*$ and $Q_3 = 3/5\ \overline{\Gamma K} = 0.2(b_1^* + b_2^*)$, where $b_1^*$ and $b_2^*$ are the reciprocal primitive vectors. In real space, the instabilities at $Q_1$ and $Q_3$ are associated with the 4×4 and $\sqrt{7}\times\sqrt{3}$ CDW supercells, suggesting that increasing the interlayer spacing induces a dimensional crossover from 3D to 2D. When the energy gains of both CDW patterns are calculated, the $\sqrt{7}\times\sqrt{3}$ configuration is more favorable. We note that the $\sqrt{7}\times\sqrt{3}$ distortion in principle can be driven primarily by the instability at $Q_3$ (corresponding to the $\sqrt{7}$ periodicity): a commensurate supercell cannot be constructed using the $\sqrt{7}$ periodicity alone, and the $\sqrt{3} \times \sqrt{7}$ superstructure naturally emerges as the minimal cell with the $\sqrt{7}$ vector as a geometric consequence in 2D (Figure S3). Thus, these results imply that the electron-phonon coupling could play an important role in the formation of CDW, particularly as the dimensionality is reduced. It has been reported that the leading instability in harmonic and anharmonic phonon calculations is switchable between $Q_1$ and $Q_3$ because of strain in monolayer VSe$_2$[41,45]. In our organic-cation-intercalated samples, the total strain in the VSe$_2$ plane is relatively weak, which favors the $\sqrt{7}\times\sqrt{3}$ order.

The insulating gap has a magnitude of approximately 115 meV, clearly surpassing the mean-field prediction for a continues CDW transition with $T_{CDW}$ ~ 150 K. Our spectroscopic measurements further reveal that this hard insulating gap is uniform across the intercalated VSe$_2$ samples (Figure S2). For example, a linecut of *dI/dV* spectra with a length of 50 nm is shown in Fig. 4A, demonstrating a rather homogeneous gap feature at the nanoscale; a similar gap value has also been checked at locations separated by hundreds of microns. We proceed to investigate the origin of this energy gap. In fact, intercalation not only expands the interlayer spacing, but also leads to substantial electron doping into the VSe$_2$ layers, as the (TBA)$^+$ cations inevitably lower the chemical valence of the vanadium ions. To determine the doping level in our (TBA)$_x$VSe$_2$ samples, we characterized a number of samples via an electron probe X-ray microanalyzer. Figure 4B presents a backscattered electron image obtained from a 1.3 mm×1.3 mm area on a freshly cleaved surface, and the high-definition element mapping analysis in this area is plotted in Fig. 4B for the major composition elements, including C, N, and V. In light of these mappings, there is no significant inhomogeneity in the distribution of the C and N atoms that stem from the intercalated organic ion (TBA)$^+$. This is consistent with the ubiquitous electronic gap in our STM measurements. Moreover, high-sensitivity quantitative analysis was performed at different locations on several samples. The average doping level *x* is determined to be 0.3 ± 0.1 via the statistical analysis in Fig. 4C, indicating that the doping concentration in the 'freestanding monolayer' could be regarded as (TBA)$_{0.3}$VSe$_2$.

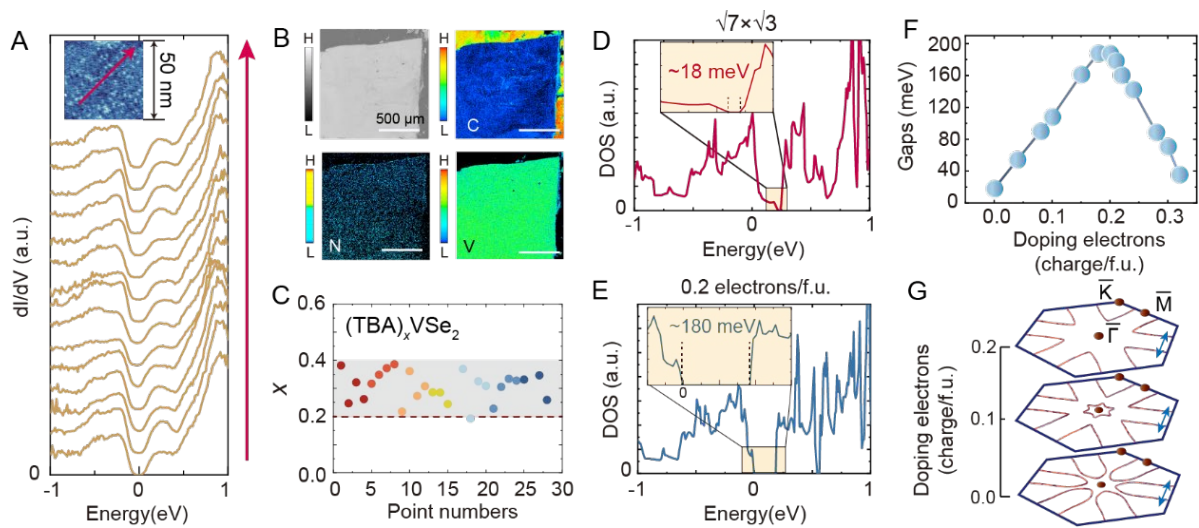

**Figure 4.** (A) Linecut of dI/dV spectra measured along the pink arrow shown in the inset (50 nm). The spectra are vertically offset for clarity ($V_S$ = 1.5 V, $I_t$ = 500 pA, $V_m$ = 10 mV). (B) Backscattered electron imaging and high-definition element (including C, N and V) mapping of the 1.3 mm×1.3 mm area on a fresh surface of $(TBA)_xVSe_2$. (C) Resulting doping level $x$ for $(TBA)_xVSe_2$. The data obtained from different samples are color-coded. (D, E) The total electronic DOS of the $\sqrt{7}\times\sqrt{3}$ CDW supercell without additional electron doping and with 0.2 electrons/f.u. doping, respectively. The insets show the zoomed-in view of the DOS near the Fermi level. (F) Gap magnitude as a function of electron doping. (G) The calculated FSs for the normal phase of $VSe_2$ for x = 0, 0.1 and 0.2, where the FS topology undergoes a Lifshitz transition. The CDW vector, $Q_{\sqrt{7}a_0}$ along $\overline{\Gamma K}$, is denoted by blue arrows.

Having determined the electron-doping level of our sample, we next studied the corresponding electronic structure. The band structures and the resulting total electronic DOS were calculated with the new $\sqrt{7}\times\sqrt{3}$ CDW supercell. For a $\sqrt{7}\times\sqrt{3}$ ordered $VSe_2$ layer without additional electron doping, the calculated DOS shows a metallic ground state with finite electron states at the Fermi level, and a small energy gap of approximately 18 meV is found for the unoccupied states (Fig. 4D). Upon electron doping, the energy gap moves toward the Fermi level (Figure S4), and more interestingly, its magnitude undergoes nonmonotonic evolution (Fig. 4F). At a doping level of approximately 0.2~0.22 electrons per formula unit (/f.u.), we find a large, hard energy gap of approximately 180 meV that spans the Fermi level (Fig. 4E), which is in good agreement with our experimental data. In contrast, we did not find any insulating gap for the undistorted monolayer $VSe_2$ with electron doping (Figure S5), highlighting the critical role of the $\sqrt{7}\times\sqrt{3}$ superlattice. Although the doping level in the calculation is smaller than that observed in our bulk measurement, we believe that this is reasonable because of the lower electron transfer of the $(TBA)^+$ cation on the fully exposed $VSe_2$ surfaces, as well as the experimental error in determining the doping concentrations. Notably, the phonon dispersions do not change significantly within these doping ranges[50]. The detailed evolution of band structures with $\sqrt{7}\times\sqrt{3}$ ordering upon electron doping is summarized in Fig. 5. The valence and conduction bands exhibit a complex transformation that does not follow rigid shift behavior.

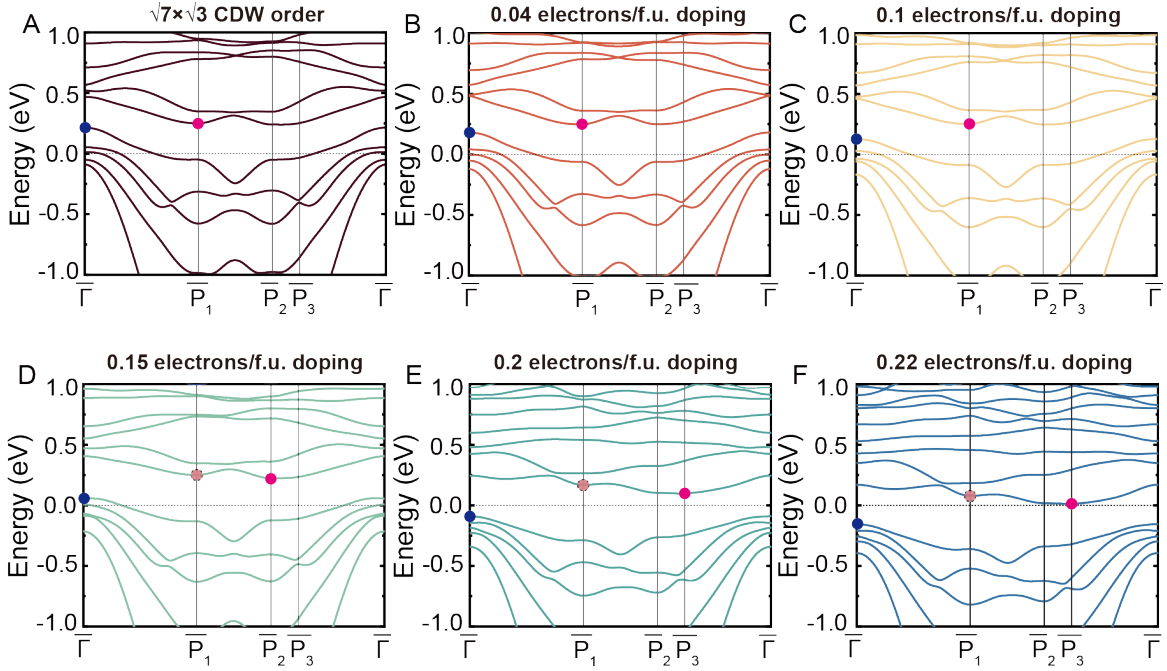

**Figure 5.** (A-F) Calculated band structures with √7×√3 ordering as a function of the doping concentration. The high symmetry paths of the band structures are plotted in Fig. 3C. The pink and dark blue circles represent the conduction band minimum and valence band maximum, respectively. The light pink circles in D-F track the location of original conduction band minimum for A-C.

Our findings suggest a heavily electron doped case of VSe$_2$ with a √7×√3 CDW order in the intercalated samples. To check the possible role of FS nesting, we calculated the band structures and Fermi surfaces for the normal 1×1 phase of VSe$_2$, as shown in Fig. 4G. Upon electron doping, the FS topology undergoes a Lifshitz transition, from elliptical electron pockets centered at the $\bar{M}$ points (case 1) to triangular hole-like pockets around the $\bar{K}$ points (case 2). However, the emergence of the √7×√3 CDW order remains robust in both cases, as evidenced by our results (case 2) and monolayer VSe$_2$ films grown on 6H-SiC (0001) (case 1)[36], despite the enlargement of the nesting vector with electron doping. Thus, we conclude that FS nesting is not the driving force for this √7×√3 order. Moreover, electron-electron interactions should also be weak in VSe$_2$, since the gap feature can be well reproduced by our calculations at the single-electron level.

Since STM is a surface probe, one may doubt that it does not reflect the bulk properties. We believe that the observation of √7×√3 CDW is intrinsic for $(TBA)_xVSe_2$ for multiple reasons. First, the √7×√3 CDW order and associated insulating gap are continuously observed on all exposed $VSe_2$ terminations. Second, the √7×√3 CDW order is consistent with the majority of studies on monolayer $VSe_2$ films. From a computational perspective, an interlayer spacing of 1.8 nm for $VSe_2$ can be regarded as a 2D system so that the surface $VSe_2$ layer is representative of the bulk. Third, the insulating gap can be reproduced via our DFT calculations when both the √7×√3 CDW order and electron doping are considered. On the other hand, electron diffraction measurements[18] have found fuzzy 3×3 CDW signals in $(TBA)_xVSe_2$, which cannot simply give rise to an insulating gap even with electron doping (Figure S5). Thus, the 3×3 order may be a possible metastable state at 130 K near the phase transition or just from the degeneration of the samples during the complicated processes of sample preparation.

Finally, we discuss the similarities and differences between Na-intercalated subsurface islands[43] on 1T-$VSe_2$ and our bulk organic-$VSe_2$ system. Despite the use of chemically distinct intercalants, the emergence of the same √7×√3 CDW indicates a common underlying mechanism for its formation by reducing the interlayer coupling. For our bulk organic-$VSe_2$ compound, the $TBA^+$ ions are intercalated into the entire crystal, which give rises to an insulating gap in the $VSe_2$ layers. However, the tunneling spectra exhibit a metallic state on the localized Na subsurface island with a small size, likely because the metallic background of bulk 1T-$VSe_2$ acts as a bath that increases the itinerary of electrons[43].

Taken together, our STM measurements and DFT calculations verified a rare case of the MIT that is associated with a √7×√3 CDW order in bulk organic-$VSe_2$ systems. The superstructure is consistent with that observed on subsurface islands formed by Na-intercalated $VSe_2$[43], further highlighting the role of interlayer coupling in the CDW phase. The large insulating gap, imaginary phonon frequencies at the characteristic wavevector and the insensitivity to the FS topology provide strong evidence for the electron-phonon-coupling origin of the √7×√3 distortion. Our results provide new insights into the rich CDW instabilities in the VSe$_2$ family and open new avenues for tailoring the exotic electronic properties of 2D layered materials.

**MATERIALS and METHODS**

**Single crystal growth of (TBA)$_x$VSe$_2$.** Two steps are involved in the synthesis of (TBA)$_x$VSe$_2$. First, single crystals of 1T-VSe$_2$ were grown via a chemical vapor transport method as described in detail elsewhere (18,51). Second, the 1T-VSe$_2$ crystals were served as the starting material for the electrochemical intercalation process to synthesize (TBA)$_x$VSe$_2$. Owing to the sensitivity of the intercalated samples to air and moisture, the sample synthesis, electrode manufacturing processes and cleavage preparation were all performed in an argon-filled glove box.

**STM/STS measurements.** The STM study was carried out on a commercial CreaTec STM. Both VSe$_2$ and (TBA)$_{0.3}$VSe$_2$ single crystals were cleaved in a vacuum better than $1\times10^{-10}$ Torr at ~77 K and were immediately inserted into the STM head. The measurements for bulk and intercalated VSe$_2$ were performed at 77 K and 4.8 K, respectively. Tungsten tips were electrochemically etched and then checked on the surface of a single crystal Au (111) before performing the measurements. Spectroscopic data were acquired via the standard lock-in technique at a frequency of 987.5 Hz under a modulation voltage of ~10 mV.

**Electronic structure calculation.** The calculations were performed via the Vienna *ab initio* simulation package (VASP)[52,53] with the projector augmented wave method[54]. The exchange-correlation interaction was parameterized by the Perdew-Burke-Ernzerhof functional within the framework of the generalized gradient approximation (GGA)[55]. The cutoff energy was set to 500 eV for all calculations. For the CDW configuration, a √7×√3 supercell was constructed, complemented by a 15 Å vacuum layer to minimize periodic boundary effects. The Γ-centered Monkhorst-Pack[56] grids of 24×24×16, 20×20×1 and 8×12×1 in the first Brillouin zone were employed for the geometric and electronic calculations of the bulk, monolayer and superstructure VSe$_2$, respectively. With the cell shape and volume fixed, all the atoms in the VSe$_2$ structure were fully optimized until the ionic forces were less than 0.001 eV/Å. To investigate the lattice dynamics, the phonon dispersion is calculated via the supercell method as implemented in the PHONOPY code[57]. We consider 4×4×3 and 8×8×1 supercells for phonon calculations of bulk and monolayer structures, respectively. Denser 6×6×4 and 3×3×1 k-meshes were used to ensure converged phonon results. Moreover, the tetrahedron method with Blöchl corrections was used for all density of states calculations.

## ASSOCIATED CONTENT

**Supporting Information**
STM topographic images of the TBA terminations; atomic resolution STM topographies of the exposed VSe$_2$ surface; geometric argument on the $\sqrt{3} \times \sqrt{7}$ superstructure; doping-dependence of the total electronic DOS in the $\sqrt{7}\times\sqrt{3}$ CDW phase; total electronic DOS with different CDW phases.

## AUTHOR INFORMATION

**Corresponding Authors**
**Zhenyu Wang** - *Department of Physics, University of Science and Technology of China, Hefei, Anhui 230026, China*; Email: zywang2@ustc.edu.cn
**Xianhui Chen** - *Department of Physics, University of Science and Technology of China, Hefei, Anhui 230026, China*; *CAS Center for Excellence in Quantum Information and Quantum Physics, Hefei, Anhui 230026, China; Collaborative Innovation Center of Advanced Microstructures, Nanjing 210093, China;* Email: chenxh@ustc.edu.cn

**Authors**
**Wanru Ma** - *Department of Physics, University of Science and Technology of China, Hefei, Anhui 230026, China*
**Ye Yang** - *Department of Physics, Hefei University of Technology, Hefei, Anhui 230009, China*
**Zuowei Liang** - *Department of Physics, University of Science and Technology of China, Hefei, Anhui 230026, China*
**Ping Wu** - *Department of Physics, University of Science and Technology of China, Hefei, Anhui 230026, China*
**Fanbao Meng** - *Department of Physics, University of Science and Technology of China, Hefei, Anhui 230026, China*

**Author Contributions**

W. M. and F. M. contribute to the materials synthesis and resistivity measurements. W. M., Z. L. and P. W. performed the STM measurements. W. M. and Y. Y. performed the DFT calculations. W. M., Z. W. and X. C. analyzed the data and wrote the manuscript.

**Notes**

The authors declare no competing financial interest.

**ACKNOWLEDGMENTS**

We thank Tao Wu and Baolei Kang for valuable discussions. This work was supported by the National Natural Science Foundation of China No. 52261135638, the Fundamental Research Funds for the Central Universities No. WK2030000111, the International Partnership Program of the Chinese Academy of Sciences (Grant No. 123GJHZ2022035MI), and the Systematic Fundamental Research Program Leveraging Major Scientific and Technological Infrastructure, Chinese Academy of Sciences under Contract No. JZHKYPT-2021-08.

# REFERENCES

(1) Wang, Q. H.; Kalantar-Zadeh, K.; Kis, A.; Coleman, J. N.; Strano, M. S. Electronics and optoelectronics of two-dimensional transition metal dichalcogenides. *Nat. Nanotech.* **2012**, *7* (11), 699–712.

(2) Chhowalla, M.; Shin, H. S.; Eda, G.; Li, L.-J.; Loh, K. P.; Zhang, H. The chemistry of two-dimensional layered transition metal dichalcogenide nanosheets. *Nat. Chem.* **2013**, *5* (4), 263–275.

(3) Manzeli, S.; Ovchinnikov, D.; Pasquier, D.; Yazyev, O. V.; Kis, A. 2D transition metal dichalcogenides. *Nat. Rev. Mater.* **2017**, *2* (8), 17033.

(4) Li, H.; Zhang, Q.; Yap, C. C. R.; Tay, B. K.; Edwin, T. H. T.; Olivier, A.; Baillargeat, D. From Bulk to Monolayer $MoS_2$: Evolution of Raman Scattering. *Adv. Funct. Mater.* **2012**, *22* (7), 1385–1390.

(5) Xi, X.; Wang, Z.; Zhao, W.; Park, J.-H.; Law, K. T.; Berger, H.; Forró, L.; Shan, J.; Mak, K. F. Ising pairing in superconducting $NbSe_2$ atomic layers. *Nat. Phys.* **2016**, *12* (2), 139–143.

(6) Wu, S. F.; Fatemi, V.; Gibson, Q. D.; Watanabe, K.; Taniguchi, T.; Cava, R. J.; Jarillo-Herrero, P. Observation of the quantum spin Hall effect up to 100 kelvin in a monolayer crystal. *Science* **2018**, *359* (6371), 76-79.

(7) Fei, Z.; Palomaki, T.; Wu, S.; Zhao, W.; Cai, X.; Sun, B.; Nguyen, P.; Finney, J.; Xu, X.; Cobden, D. H. Edge conduction in monolayer $WTe_2$. *Nat. Phys.* **2017**, *13* (7), 677–682.

(8) Tang, S.; Zhang, C.; Wong, D.; Pedramrazi, Z.; Tsai, H.-Z.; Jia, C.; Moritz, B.; Claassen, M.; Ryu, H.; Kahn, S.; Jiang, J.; Yan, H.; Hashimoto, M.; Lu, D.; Moore, R. G.; Hwang, C.-C.; Hwang, C.; Hussain, Z.; Chen, Y.; Ugeda, M. M.; Liu, Z.; Xie, X.; Devereaux, T. P.; Crommie, M. F.; Mo, S.-K.; Shen, Z.-X. Quantum spin Hall state in monolayer 1T'-$WTe_2$. *Nat. Phys.* **2017**, *13* (7), 683-687.

(9) Zhao, W.; Runburg, E.; Fei, Z.; Mutch, J.; Malinowski, P.; Sun, B.; Huang, X.; Pesin, D.; Cui, Y.-T.; Xu, X.; Chu, J.-H.; Cobden, D. H. Determination of the Spin Axis in Quantum Spin Hall Insulator Candidate Monolayer $WTe_2$. *Phys. Rev. X* **2021**, *11* (4), 041034.

(10) Xi, X.; Zhao, L.; Wang, Z.; Berger, H.; Forró, L.; Shan, J.; Mak, K. F. Strongly enhanced charge-density-wave order in monolayer $NbSe_2$. *Nat. Nanotech.* **2015**, *10* (9), 765–769.

(11) Chen, P.; Chan, Y.-H.; Fang, X.-Y.; Zhang, Y.; Chou, M. Y.; Mo, S.-K.; Hussain, Z.; Fedorov, A.-V.; Chiang, T.-C. Charge density wave transition in single-layer titanium diselenide. *Nat. Commun.* **2015**, *6* (1), 8943.

(12) Chen, P.; Pai, W. W.; Chan, Y.-H.; Takayama, A.; Xu, C.-Z.; Karn, A.; Hasegawa, S.; Chou, M. Y.; Mo, S.-K.; Fedorov, A.-V.; Chiang, T.-C. Emergence of charge density waves and a pseudogap in single-layer $TiTe_2$. *Nat. Commun.* **2017**, *8* (1), 516.

(13) Yu, Y.; Yang, F.; Lu, X. F.; Yan, Y. J.; Cho, Y.-H.; Ma, L.; Niu, X.; Kim, S.; Son, Y.-W.; Feng, D.; Li,

## TOC Graphic

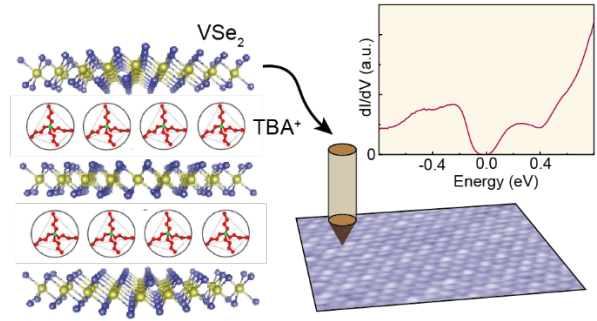